\documentclass[11pt,oneside]{article} 	

\usepackage{graphicx}
\usepackage{amssymb,amsmath}
\usepackage{epsfig}
\usepackage{txfonts}

\usepackage{amsmath,amssymb,amsfonts}
\usepackage{mathrsfs}
\usepackage{empheq}

\usepackage{color, soul}
\usepackage[unicode,breaklinks]{hyperref}

\usepackage[unicode]{hyperref}
\hypersetup{
	pdftitle={review},
	pdfauthor={Yu, Li},
	pdfsubject={populationdynamics},
	colorlinks=true,
	linkcolor=blue,
	citecolor=blue,
	filecolor=black,
	urlcolor=blue
}

\def\url#1{}

\usepackage{geometry}
\usepackage{natbib}
\usepackage[parfill]{parskip} 
\usepackage{graphicx}
\usepackage{amsmath,amssymb}
\usepackage[usenames,dvipsnames]{xcolor} 
\usepackage{textcomp}
\usepackage{float}
\usepackage{caption} 
\usepackage{authblk}

\usepackage{soul} 
\usepackage[colorinlistoftodos]{todonotes} 

\usepackage{lineno}

\title{Applications of WKB and Fokker-Planck methods in analyzing population extinction driven by weak demographic fluctuations}

\author[1]{Xiaoquan Yu}
\author[2]{Xiang-Yi Li* \{xiangyi.li@ieu.uzh.ch\}}
\affil[1]{Department of Physics, Centre for Quantum Science, and Dodd-Walls Centre for Photonic and Quantum Technologies, University of Otago, Dunedin 9054, New Zealand}
\affil[2]{Department of Evolutionary Biology and Environmental Studies, University of Zurich, Winterthurerstrasse 190, 8057 Zurich, Switzerland}

\date{\today}

\begin{document}
\maketitle
\begin{abstract}
In large but finite populations, weak demographic stochasticity due to random birth and death events can lead to population extinction.
The process is analogous to the escaping problem of trapped particles under random forces. 
Methods widely used in studying such physical systems, for instance, Wentzel-Kramers-Brillouin (WKB) and Fokker-Planck methods, can be applied to solve similar biological problems. 
In this article, we comparatively analyse applications of WKB and Fokker-Planck methods to some typical stochastic population dynamical models, including the logistic growth, endemic SIR, predator-prey, and competitive Lotka-Volterra models. 
The mean extinction time strongly depends on the nature of the corresponding deterministic fixed point(s).
For different types of fixed points,  the extinction can be driven either by rare events or typical Gaussian fluctuations.  
In the former case,  the large deviation function that governs the  distribution of rare events  can be well-approximated by the WKB method in the weak noise limit.  
In the later case, the simpler Fokker-Planck approximation approach is also appropriate. 
\end{abstract}
\newpage
\tableofcontents

\section{Introduction}

The extinction of local populations can happen frequently in nature, particularly in small and fragmented habitats due to various causes, including genetic deterioration, over-harvesting, climate change, and environmental catastrophes.
Even in the absence of all other causes, the finiteness of population size and the resultant demographic stochasticity will eventually drive any isolated population to extinction. 
Therefore, the expected time until population extinction due to demographic stochasticity alone provides a baseline scenario estimation for the long-term viability of the population.
It is closely related to the concept of minimal viable population size, and is of great importance to the conservation of species and global biodiversity \citep{shaffer:BS:1981,traill:BC:2007}.

The study of population extinction due to demographic stochasticity is a long-standing yet rapidly advancing topic of research, with Francis Galton's famous problem of the extinction of family names already proposed in 1873 (for reviews of the history see \cite{kendall:JMS:1966}).
In the last decades, new mathematical tools have been developed to analyse stochastic population dynamics.
A number of such tools, such as the Fokker-Planck approximation and the Wentzel-Kramers-Brillouin (WKB) approximation methods, were originally developed for solving problems in statistical mechanics and quantum mechanics.
We can take advantage of analogies between biological systems and corresponding physical systems (e.g., the extinction of population from a steady state driven by weak noise is very similar to the escaping problem of particles in a trapping potential~\citep{DykmanJCP1994}), and apply methods developed for tackling physical problems to answering biological questions.

In this paper, we provide a pedagogical comparative study of the WKB and Fokker-Planck approximation methods in analyzing population extinction from a stable state driven by weak demographic fluctuations.
We examine some widely-used stochastic models of population extinction as examples, and show that the nature of the stable states in the mean-field level determines the behaviour of the mean extinction time.
In systems with an attracting fixed point or limit cycle, extinction is caused by rare events, the 
WKB method is a natural approach.
For systems with marginally stable states, since extinction is driven by typical Gaussian fluctuations, the Fokker-Planck approximation is also valid. 

\color{black}
\section{Extinction time of populations formed by a single species}
\label{singlespecies}

\subsection{The deterministic logistic growth model}
One of the most widely applied population growth model of a single species is the logistic growth model, or the Verhulst model~\citep{verhulst:Quetelet:1838}.
This model has been extensively used in modelling the saturation of population size due to resource limitations \citep{murray:book:2007, mcelreath:book:2008, haefner:book:2012}, and formed the basis for several extended models that predict more accurately the population growth in real biological systems, such as the Gompertz, Richards, Schnute, and Stannard models (for a review, see \cite{tsoularis:MB:2002}).

The classic logistic model takes the form

\begin{align}
\frac{dn}{dt}=r n \left(1-\frac{n}{K}\right),
\label{Eq.LogisticGrowth}
\end{align}

where $n$ represents population size, the positive constant $r$ defines the growth rate and $K$ is the carrying capacity.
The unimpeded growth rate is modeled by the first term $rn$ and the second term captures the competition for resources, such as food or living space. The solution to the equation has the form of a logistic function

\begin{align}
n(t) = \frac{K n_0 e^{rt}}{K + n_0 \left( e^{rt} - 1\right)}, 
\end{align}

where $n_{0}$ is the initial population size.  
Note that $\displaystyle \lim _{t\to \infty }n(t)=K$, and this limit is asymptotically reached as long as the initial population size is positive, and the extinction of the population will never happen.

\subsection{Population dynamics under demographic stochasticity}
When the typical size of the population  is very large ($1/K \ll 1$), fluctuations in the observed number of individuals are typically small.
In this case, the deterministic logistic growth model generally provides a good approximation to the population dynamics by predicting that the population will evolve towards and then persists at the stable stationary state where $n=K$.
However, in the presence of the demographic noise, occasional large fluctuations can still induce extinction, making the stable states in the deterministic level metastable.  
In any finite population, extinction
will occur as $t\rightarrow\infty$ with unit probability.  

In an established population under logistic growth with a large carrying capacity, the population size fluctuates around $K$ due to random birth and death events, and typically the fluctuation is small in the large $K$ limit.
But from time to time, a rare large fluctuation can happen, and it may lead to the extinction of the population. 
In such situations, it is interesting and often biologically important to determine the most probable paths and the mean extinction time, starting from the stable population size. 
A rigorous approach for solving these problems in the weak noise limit is the large deviation theory~\citep{touchette2009large}.
We use the logistic growth model to illustrate the main idea. 

Let the function $T(n\rightarrow m)$ represent the probability of the transition $n\rightarrow m$ per unit time. 
For the logistic model $T(n\rightarrow n+1)=\lambda_n=Bn$ describes the birth rate of the popualtion, where $B$ is the per capita growth rate, and $T(n \rightarrow n-1)=\mu_n=n+B n^2/K$ describes the death rate of the population, in which the first term represents spontaneous death, and the second term represents death caused by competition.
The function $P(n,t)$ is the probability density for the system to be in the state with the population of $n$ at the time $t$, obeying a Master equation 

\begin{align}
\label{master}
\frac{d P(n,t)}{ d t}&=\sum_{m} \left[T(m \rightarrow n,t) P(m,t)-T(n \rightarrow m,t) P(n,t)\right] \nonumber \\
 &=\mu_{n+1}P(n+1,t)+\lambda_{n-1}P(n-1,t)-(\mu_{n}+\lambda_{n})P(n,t).
\end{align}

The initial condition $P(n,t=t_0)=\delta_{n,n(0)}$. Since $n=0$ is an absorbing state, for $m>0$, $T(0 \rightarrow m)=0$, we have 

\begin{align}
\frac{P(n=0,t)}{dt}=\sum_{m>0} T(m\rightarrow 0) P(m,t).
\end{align}

The average population size $\overline{n}=\sum_{n} P(n,t) n$ satisfies a deterministic averaged (mean-field) rate equation

\begin{align}
\label{avergeVerhulst}
\frac{d\overline{n}}{dt}=(B-1)\overline{n}-B\frac{\overline{n}^2}{K},
\end{align}
where we neglect the number fluctuation, namely, $\overline{n^2}=\overline{n}^2$ (mean-field).
Now we have derived the stochastic version of the logistic growth function, corresponding to Eq.~\eqref{Eq.LogisticGrowth}.
Eq.~\eqref{avergeVerhulst} has two fixed points: 
an attracting fixed point $\overline{n}_{\rm s}=(B-1)K/B$, provided $B>1$; and a repelling fixed point $\overline{n}_{\rm e}=0$ (extinction point). 
In the presence of noise there is a quasi-stationary state for $B>1$, in which the population fluctuates near $n_{\rm s}$. 
However, the system eventually is going to reach $n=n_{\rm e}=0$ driven by rare events, where extinction happens. 
It is then important to estimate the extinction time. 

The commonly used methods for estimating the time until extinction include the Fokker-Planck approximation (also called \textit{diffusion approximation} in population genetics literature), and the Wentzel-Kramers-Brillouin (WKB) method.
The former has a long history of application in studying biological population dynamics, going back to \cite{fisher:PRSE:1922}, and was greatly promoted since the seminal work of \cite{kimura:JAP:1964}.
Nowadays it has become an indispensable topic in population genetics textbooks \citep{ewens:book:2004, Svirezhev:book:2012}.
But despite its honourable place in mathematical biology, the application of Fokker-Planck approximation is restricted to systems where the extinction is driven by typical Gaussian fluctuations (such as genetic drift), characterised by frequent but small jumps \citep{gardiner1985handbook}.
The WKB method was introduced into biology much later (most works are published only in the last two decades), yet it has been gaining popularity steadily, as it generally  provides more accurate predictions of the mean extinction time if the extinction is driven by rare events, and can be applied under much broader conditions.

In the following we will first introduce the more general WKB method and then the classic Fokker-Planck approximation, in order to facilitate the comparison of the two methods later on.

\subsubsection{Wentzel-Kramers-Brillouin (WKB) method}

The Wentzel-Kramers-Brillouin (WKB) method  was named after the three physicists Gregor Wentzel~\citep{wentzel1926}, Hendrik Kramers~\citep{kramers1926} and L{\'e}on Brillouin~\citep{brillouin1926}.  
It provides a systematic and controllable  approximating method to calculate the mean extinction time in the small fluctuations limit. 
And it has been applied widely in studying different extinction problems, such as large fluctuations in numbers of molecules in chemical reactions \citep{DykmanJCP1994}, the fixation of a strategy in evolutionary games \citep{Black2012prl}, and the extinction of epidemics
~\citep{Chen2017JSM} .

In a finite population under logistic growth, once the stationary state is reached, the population size fluctuates around the metastable attractor $\bar{n}_{\rm s}$.
The characteristic scale of the fluctuations is of the order of $1/\sqrt{K}$ (Central Limit Theorem). 
However, occasionally much larger fluctuations also happen that take the system far from the stable state~\citep{DykmanJCP1994}. 
Such large fluctuations are rare events, and their probabilities form the tails of the quasi-stationary population state distribution. 
The mean extinction time $\tau$ (mean time to reach the absorbing state $n_{\rm e}=0$) is determined by this quasi-stationary distribution according to the Fermi's golden rule~\citep{FermiGolden}

\begin{align}
\label{decaytime}
\tau^{-1}=\sum_{n>0} T(n\rightarrow 0) P_{\rm st}(n),
\end{align}

where the stationary distribution $P_{\rm st}(n)$ satisfies 

\begin{align}
\label{statinarymaster}
0=\sum_{m} \left[T(m \rightarrow n,t) P_{\rm st}(m)-T(n \rightarrow m,t) P_{\rm st}(n)\right]. 
\end{align}

In terms of the rescaled population size $x=n/K=n \epsilon$ with $\epsilon=1/K$,
$\lambda (x)=\lambda_n/K=Bx$, and $\mu (x)=\mu_n/K=x+Bx^2$.
We look for the solution of Eq.~\eqref{statinarymaster} by proposing a large deviation form of the stationary distribution 

\begin{align}
\label{LD}
P_{\rm st}(x) = C \exp\left(-\mathcal S_{\epsilon}/\epsilon\right) 
\end{align} 
with the  WKB ansatz: $\mathcal S_{\epsilon}=\sum^{\infty}_{i=0} \epsilon^{i} \mathcal S_{i}$. 
Here $\epsilon $ characterises the noise level, and at the weak-noise limit, $\epsilon \rightarrow 0$. 
An asymptotic expansion in small $\epsilon$ corresponds to a semiclassical approximation. 
In both quantum mechanics and statistical mechanics this is also known as a WKB expansion. 
In the former case, $\epsilon$ is the Planck constant $\hbar$, characterising quantum fluctuations; and in the later case, $\epsilon$ is the temperature, characterising thermal fluctuations. 
In stochastic population dynamics,  meanwhile, the small parameter $\epsilon$ is $1/K$, characterising population size fluctuations.
 
Plugging Eq.~\eqref{LD} into Eq.~\eqref{statinarymaster} and expanding $\mathcal S_{\epsilon}$ to $\mathcal{O}(\epsilon)$, we obtain  

\begin{align}
\mathcal S_0(x)=\int^{x} p(x') \ dx',  \quad  \mathcal S_1(x)=\frac{1}{2} \ln [\mu(x)\lambda(x)] 
\end{align} 

where $p(x)=\ln\left[\mu(x)/\lambda(x)\right]=\ln\left[(1+Bx)/B\right]$. 
It is possible to construct an effective Hamiltonian such that the solution describes an optimal
path which represents the ground (lowest-energy) state of the effective Hamiltonian:

\begin{align}
H(x,p)=\lambda(x)(e^{p}-1)+\mu(x)(e^{-p}-1),
\end{align} 

where the canonical momentum $p=\partial \mathcal S_0/ \partial x$.

We hence obtain the stationary distribution  

\begin{align}
\label{staticdistributionsingle}
P_{\rm st}(x)=\frac{B-1}{\sqrt{2\pi K B x^2(1+Bx)}} e^{-K \mathcal S_0(x)},
\end{align}
where 

\begin{align}
\label{action}
\mathcal S_0(x)=1-B^{-1}-x+(x+B^{-1})\ln (x+B^{-1}).
\end{align}

The leading-order WKB action $\mathcal S_0$ describes an effective exponential barrier to extinction and the prefactor in Eq.~\eqref{staticdistributionsingle} is proportional to $e^{-\mathcal S_1(x)}$.

Using Eq.~\eqref{decaytime} we obtain the mean extinction time for the logistic growth model~\citep{Assafpre2010} for $1/K \ll x \ll 1/\sqrt{K}$,
\begin{align}
\label{extinctiontime}
\tau=\sqrt{\frac{2 \pi B}{N}}\frac{1}{(B-1)^2} e^{K\mathcal S_{0}(0)},
\end{align}
which is exponentially large in $K$. The analytical result of the mean extinction time Eq~.\eqref{extinctiontime} shows excellent agreement with Monte Carlo simulations~\citep{WKBAssaf}.

In this section we derived the mean extinction time of a population under logistic growth in a pedagogical way, for illustrating the basic concepts and techniques of the WKB method.
For more applications of the WKB method in single species stochastic population models, \cite{ovaskainen:TREE:2010} provide an excellent overview.
A recent review of~\cite{WKBAssaf} includes various applications of the WKB method in multi-species population dynamics.
On the technical aspect, an introduction to the path integral representation of master equations can be found in \cite{weber2017master}.

\subsubsection{Fokker-Planck approximation method}
The master equation, the exact formulation of the stochastic population dynamics, is generally difficult to solve. The WKB method provides a systematic and controllable way to approximately solve the stationary master equation by utilising the small parameter $\epsilon=1/K$.
Another way of approximately solving the master equation is to start from a formal Kramers-Moyal expansion: 

\begin{align}
\label{KMexpansion}
\frac{\partial P(X,t)}{\partial t}=\sum^{\infty}_{m=1}\frac{(-1)^{m}}{m!}\frac{\partial^{m}}{
	\partial X^{m} }\left[a_m(X,t)P(X,t)\right],
\end{align}

where 

\begin{align}
a_{m}(X,t)=\int dY (Y-X)^{m} T(X\rightarrow Y).
\end{align}

Pawula Theorem states that the expansion in Eq.~\eqref{KMexpansion} may stop either up to the second term, or must contain an infinite number of terms. 
If the expansion stops after the second term, it is called the Fokker-Planck equation~\citep{risken1996fokker}.  
Van-Kampen made the Kramers-Moyal expansion controllable by introducing a small parameter that is the inverse of a system size $\Omega^{-1}$~\citep{gardiner1985handbook}.
In the context of population dynamics governed by the logistic growth function, $\Omega$ corresponds to the carrying capacity $K$, and the random variable $X$ in Eq.~\eqref{KMexpansion} corresponds to the population size $n$.
Since we use the example of logistic growth through out section 2, we will trade generality for consistency and use $K$ and $n$ in the following. 
In terms of the scaled variable $x=n/K$, $a_m \sim K^{1-m/2}$, the Kramers-Moyal expansion will stop at the second term when $K$ is large, and the system reduces to the Fokker-Planck equation.
However, the Van-Kampen system size expansion should be used with caution. 
It may be valid only when $x$ is in the vicinity of its fixed point. 
For the rare events driven by large fluctuations,  the Fokker-Planck approximation may yield large errors. 

For the logistic growth model, the system size is characterised by the carrying capacity $K$. 
In terms of rescaled variable $x=n/K$, the master equation~\eqref{master} becomes 

\begin{align}
\label{Master}
\frac{d P(x,t)}{ d t}
&=K\mu (x+\delta x)P(x+\delta x,t)+K\lambda(x-\delta x)P(x-\delta x,t)-K(\mu(x)+\lambda(x))P(x,t),
\end{align}

where $\delta x=1/K$.
Expanding Eq.~\eqref{Master} to $(\delta x)^2$, we obtain the Fokker-Planck equation 

\begin{align}
\label{FP}
\frac{d P(x,t)}{ d t}
& = \frac{1}{2K}\frac{\partial^2 (g^2 P)}{\partial x^2} -\frac{\partial (f P)}{\partial x},
\end{align}

where $g^2=\lambda+\mu=(B+1)x+Bx^2$ and $f=\lambda-\mu=(B-1)x-Bx^2$.
In population genetics literature, the first term is often attributed to the effect of genetic drift, and the second term is attributed to directional selection \citep{kimura:JAP:1964, ewens:book:2004}. 
A diffusive process described by a Fokker-Planck equation, can be equivalently described by a corresponding Langevin type stochastic differential equation~\citep{gardiner1985handbook}. For Eq.~\eqref{FP}, the corresponding stochastic differential equation reads

\begin{align}
\label{Langevin}
d x =f(x,t)+K^{-1/2}g(x,t) dW(t),
\end{align}

where $W(t)$ a Wiener process with $\langle W(t) W(t') \rangle=\delta(t-t')$. Note that higher correlations functions of $W(t)$ vanish, reflecting that the stochastic process is diffusive and being consistent with the Fokker-Planck equation. 

The stationary distribution of Eq.~\eqref{FP} reads~\citep{gardiner1985handbook} 

\begin{align}
P_{\rm st}(x)\propto e^{-K \mathcal{S}_{\rm FP}(x)},
\end{align}

where $0<x<x_{\rm s}=\bar{n}_{\rm s}/K$ and the effective potential  

\begin{align}
\label{actionFP}
\mathcal{S}_{\rm FP}(x)=\int^{x_{\rm s}}_{x} dy \frac{2f(y)}{g^2(y)}=2\left[x-1+B^{-1}-2\ln \left(\frac{1+B+Bx}{2B}\right)\right].
\end{align}

\begin{figure}[!t]
	\center
	\includegraphics[width=3.0in]{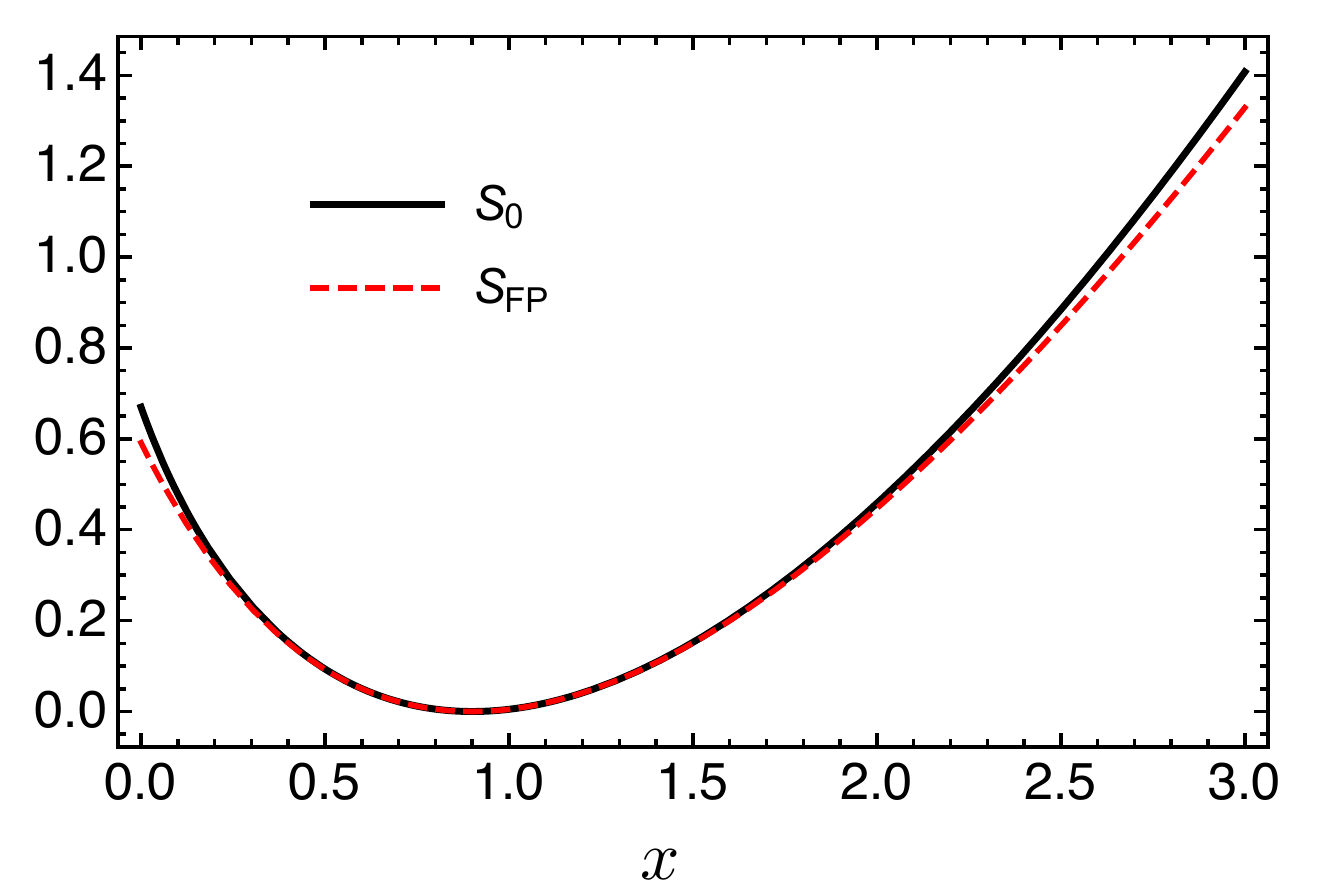}
	\caption{Comparison between $\mathcal{S}_0(x)$ in Eq.~\eqref{action}   and $\mathcal{S}_{\rm FP}(x)$ in Eq.~\eqref{actionFP} for $B=10$.} \label{compare}
\end{figure}

In the vicinity of the stable point (attracting fixed point in the deterministic level)  $x_{\rm s}=(B-1)/B$,
$\mathcal{S}_{0}(x)\simeq \mathcal{S}_{\rm FP}(x)\simeq (x-x_{\rm s})^2\ll 1$, leading to the Gaussian fluctuation.
A comparison between $\mathcal{S}_0(x)$ and $\mathcal{S}_{\rm FP}(x)$ for different  $x$ values is shown in Fig.~\ref{compare}.
Near the stable fixed point, fluctuations are Gaussian, and hence the stochastic processes can be well-approximated by the Fokker-Planck equation.  
But if we are interested in rare events driven by large fluctuations, for example the extinction event, the Fokker-Planck approximation becomes invalid. As is shown in the previous section, the mean extinction time is determined by the effective potential $\mathcal S_{\text{FP}}$ at $x=0$ which is far from $x_{\rm s}$ for $B\neq 1$. 

Compare the effective potential given by the WKB approximation

\begin{align}
\mathcal{S}_0(0)=1-B^{-1}+B^{-1}\ln B^{-1},
\end{align} 

and the corresponding result given by Fokker-Planck approximation

\begin{align}
\mathcal{S}_{\rm FP}(0)=2\left\{-1+B^{-1}-2\ln\left[(1+B)/2B\right]\right\},
\end{align} 
we can see that although Fokker-Planck approximation predicts the correct behaviour of the mean extinction time, namely, $\tau \sim e^{c K}$, it yields an error that is exponentially large in $K$~\citep{doering:MMS:2005,bressloff2014path}.
Only in the special case when $B\rightarrow 1$, $\mathcal{S}_0(0)-\mathcal{S}_{\rm FP}(0)=o ((B-1)^2)$ can be neglected. 
In this limit, $x_{\rm s} \rightarrow 0$, and hence the extinction is a typical event driven by Gaussian fluctuations. 
In summary, the Fokker-Planck approximation is valid only under the special case if $B\rightarrow 1$ and extinction is driven by typical Gaussian fluctuations, but for $B>1$, the extinction is caused by rare events, and the Fokker-Planck approximation fails to give accurate estimations of the mean extinction time.

The difference in the range of application between the WKB method and the Fokker-Planck method arises from the fundamental difference between the Master equation and the Fokker-Planck equation. 
A diffusion process characterised by the Fokker-Planck equation can always be approximated by a jump process described by the Master equation, while the reverse is true only under the conditions that the jumps must be frequent and the step sizes of such jumps must be small comparing to the time and length scales of observables~\citep{gardiner1985handbook}.

\section{Extinction time of populations of two interacting species}
In populations of two interacting species (e.g. predator and prey) or two different types of individuals (e.g. susceptible and infected), the equilibrium state predicted by the deterministic rate functions can either be a stable fixed point, a stable limit cycle, marginal stable cycles, or no attractor at all.
In general,  for an attracting fixed point or a stable limit cycle, the extinction from a stable quasi-stationary coexistence state is a rare event driven by large fluctuations, and the mean extinction time will be exponentially large in population size. 
In this situation the Fokker-Planck approximation is invalid, whereas the WKB approximation method can provide fully controlled weak noise expansion. 
But if the coexistence state is marginally stable, then the extinction event is a diffusion process driven by typical fluctuations but not a jump. 
In this case the Fokker-Planck approximation is also valid and the mean extinction time grows algebraically with the initial population size. 
We discuss the different cases separately in the following.      

\subsection{Extinction from an attracting fixed point}
As an example of multi-species stochastic systems with an attracting fixed point, we consider the endemic SIR model.
The SIR model describes the spread of a disease in a population, with susceptible ($S$), infected ($I$) and recovered ($R$) individuals.
Assuming that $N$ is the total population size at equilibrium, individuals are born (as susceptible) at rate $\mu N$. 
Susceptible, infected, and recovered individuals die at rates $\mu S$, $\mu_I I$, and $\mu_R R$, respectively.
Susceptible individuals become infected at rate $(\beta/N)SI$, and infected individuals recover at rate $\gamma I$.
The corresponding deterministic rate equations for the SIR model are

\begin{align}
\begin{split}
\label{SI}
\frac{dS}{dt}&=\mu N-\mu S-(\beta/N) S I,  \\
\frac{dI}{dt}&=-\mu_I I - \gamma I +(\beta/N) S I, \\
\frac{dR}{dt}&=-\mu_R R + \gamma I.
\end{split}
\end{align}

According to this formulation, the $R$ individuals obtain lifelong immunity and will never become $S$ or $I$ again, its dynamics is thus decoupled from that of the other two subpopulations.
For simplicity, we will ignore the $R$ individuals, and focus on the population dynamics of only $S$ and $I$ individuals.
By setting $\mu_I+\gamma=\Gamma$, which measures the effective death rate of the infected, we obtain the corresponding SI model:

\begin{align}
\begin{split}
\label{SI}
\frac{dS}{dt}&=\mu N-\mu S-(\beta/N) S I,  \\
\frac{d I}{dt}&=-\Gamma I +(\beta/N) S I.
\end{split}
\end{align}

For a sufficiently high infection rate, $\beta >\Gamma$,
there is an attracting fixed point $\bar{S}=N \Gamma  / \beta$, $\bar{I}=\mu (\beta-\Gamma)N/(\beta \Gamma)$,
corresponding to an endemic state, and an unstable
fixed point $\bar{S}=N$, $\bar{I}=0$, describing an uninfected steady-state population.

Accounting for the demographic stochasticity and random 
contacts between the susceptible and infected, the master equation for the probability $P(n,m,t)$ of finding $n$ susceptible and $m$ infected individuals at time $t$ reads

\begin{align}
\label{SImaster}
\frac{dP(n,m,t)}{dt}&=\mu \left[N(P(n-1,m)-P(n,m))+(n+1)P(n+1,m)-nP(n,m)\right]\nonumber\\
&+\Gamma\left[(m+1)P(n,m+1)-mP(n,m)\right]\nonumber\\ &+(\beta/N)\left[(n+1)(m-1)P(n+1,m-1)-nmP(n,m)\right].
\end{align}

In a finite population, the extinction of the disease, starting from the quasi-stationary endemic state, occurs within finite time due to rare events. 
It therefore is interesting to find out the mean time it takes for the $I$ subpopulation to go extinct. 
For weak fluctuations ($1/N \ll 1$), a long lived quasi-stationary distribution has a Gaussian peak around the stable state of the deterministic model. 
The Fokker-Planck approximation to the master equation can accurately describe small deviations from the stable state, but it fails to describe the probability of large fluctuations.

In section~\ref{singlespecies} we discussed the WKB approximation used directly to the quasi-stationary distribution that solves the stationary master equation. \citet{elgart2004rare} proposed a method approximating the evolution equation for the probability generating function. The generating function associated with the probability distribution is defined as  

\begin{align}
G(p_{S},p_{I},t)=\sum_{n,m}p^n_{S}p^m_{I}P(n,m,t).
\end{align}	

Using ansatz $G(p_{S},p_{I},t)=\exp[-S_{\epsilon}(p_{S},p_{I},t)/\epsilon]$ with $S_{\epsilon}(p_{S},p_{I},t)=\sum_{i=0}\epsilon^{i}S_{i}$ and $\epsilon=1/N$, to the leading order in $\epsilon$, one obtains the  
Hamilton-Jacobi equation $\partial_t \mathcal{S}_0+H=0$, where $H$ is the effective classical Hamiltonian~\citep{Alex2008pre}:

\begin{align}
\label{SIHamiltonain}
H=\mu(p_{S}-1)(N-S)-\Gamma(p_{I}-1)I-(\beta/N)(p_{S}-p_{I})p_{I} SI.
\end{align}

\begin{figure}[!t]
	\includegraphics[width=2.3in]{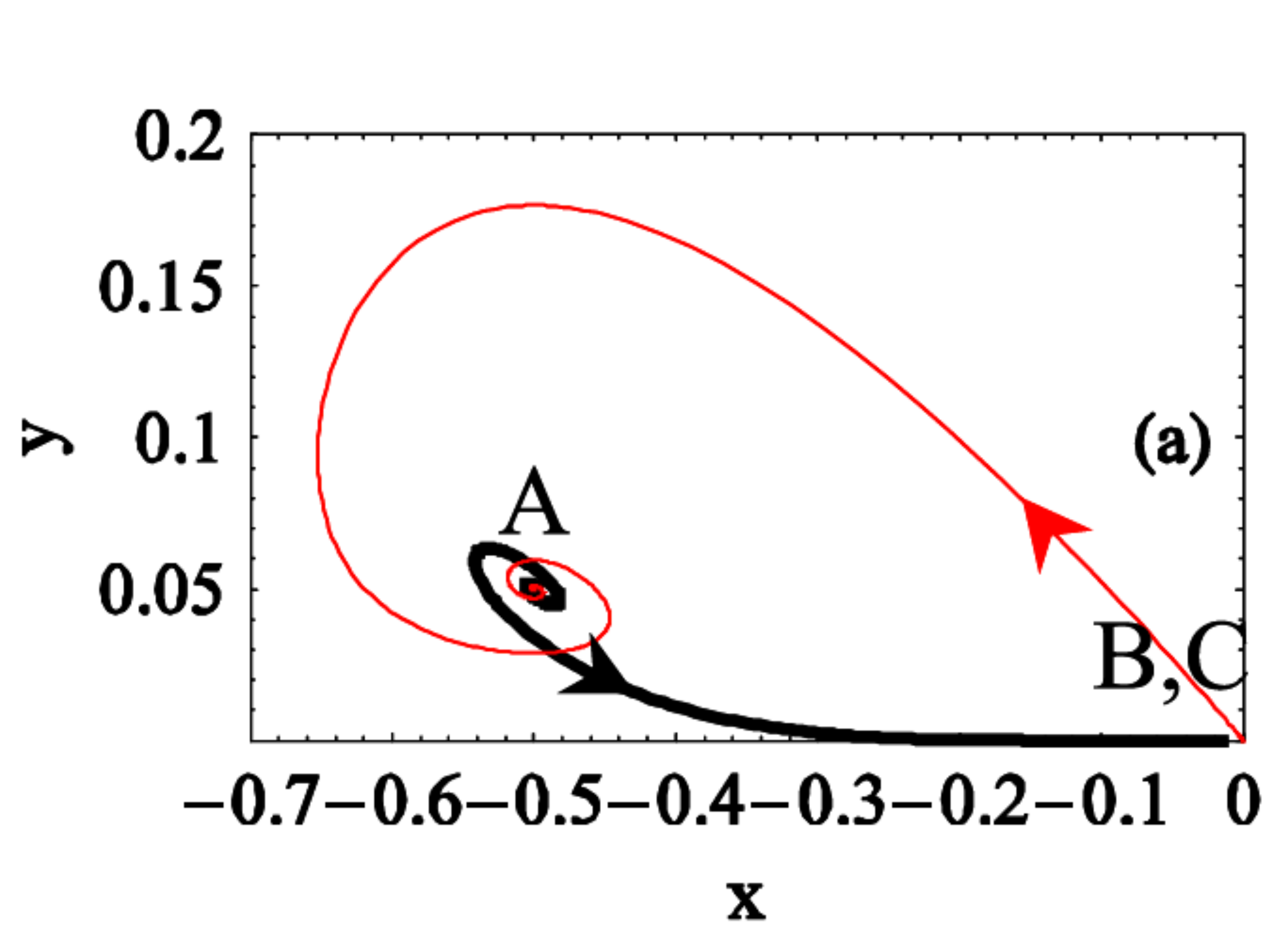}
	\quad\quad\quad\quad\quad
	\includegraphics[width=2.1in]{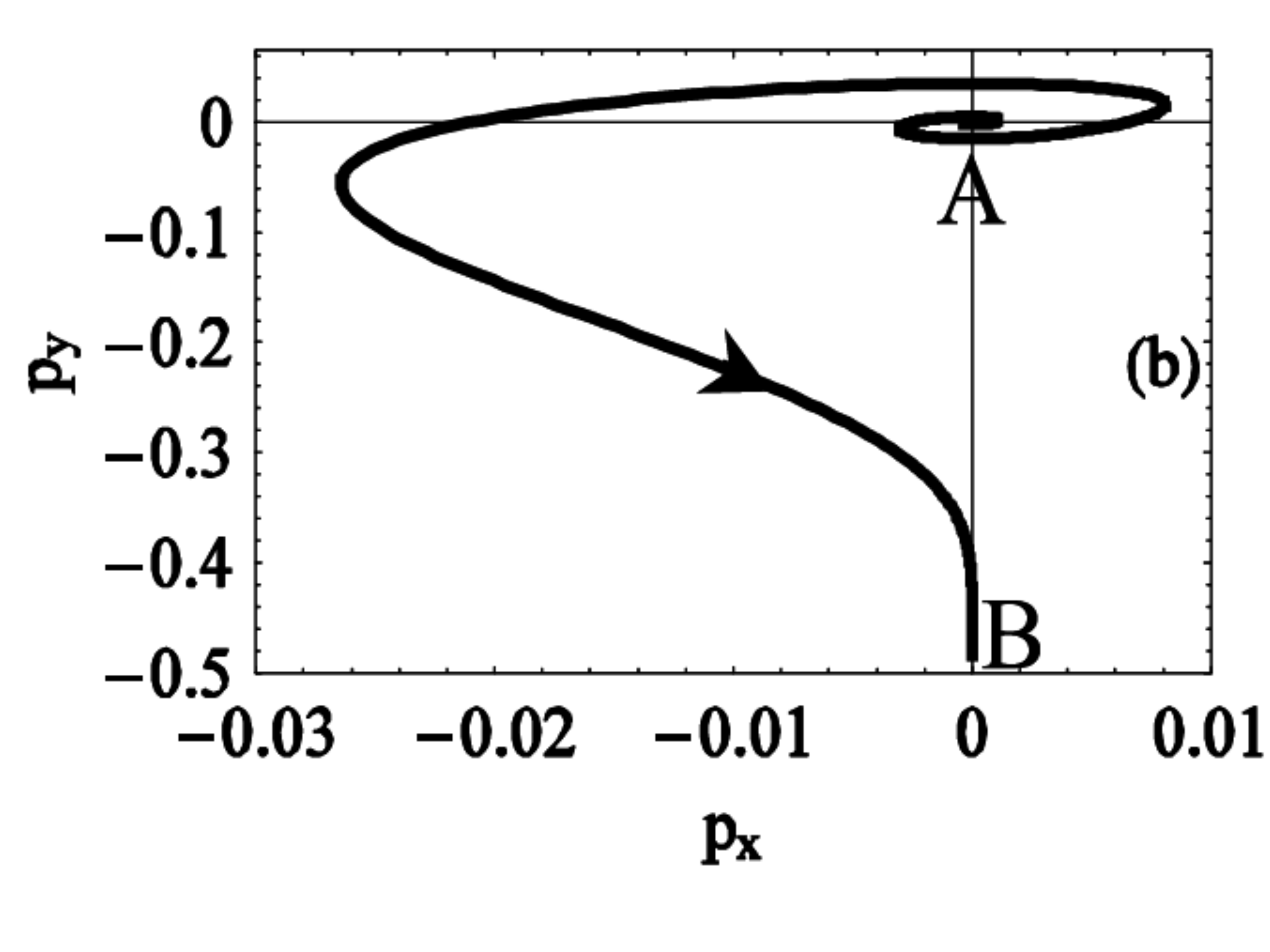}
	\caption{(a) Projection of the optimal path on the ($x$,$y$) plane (thick black line) and the mean-field trajectory ($p_x=p_y=0$) describing an epidemic outbreak (thin red line). (b) Projection of the optimal path on the ($p_x$, $p_y$) plane. $x=S/N-1$, $y=I/N$; $K=20$ and $\delta \equiv 1-\Gamma/\beta=0.5$ \citep{Alex2008pre}. Permission for reuse obtained from the publisher.} \label{SIextinction}
\end{figure}

The meanings of $p_{S}$ and $p_{I}$ are clear now. They are the canonical momenta of $S$ and $I$ respectively, and $S=-\partial_{p_{S}} \mathcal{S}_0$ and $I=-\partial_{p_{I}}\mathcal{S}_0$. 
The phase space defined by the Hamiltonian in Eq.~\eqref{SIHamiltonain} provides an important tool to study the extinction dynamics. 
Demographic stochasticity that induces the extinction of the disease proceeds along the optimal path: a particular trajectory in the phase space. 
All the mean-field trajectories, described by Eqs.~\eqref{SI} are located in the zero energy $H=0$ plane $p_S=p_I=1$.  
As illustrated in Fig.~\ref{SIextinction},
the attracting fixed point of the mean-field theory
becomes a hyperbolic point $A=[\bar{S},\bar{I},1,1]$ in this phase space. 
There are two more zero-energy fixed points in the system: the point $C=[N,0,1,1]$ that is present in the mean-field description, and the emergent fixed point $B=[N,0,1,\Gamma/\beta]$ due to the presence of fluctuations. 
Both of them are hyperbolic and describe extinction of the disease. 

The optimal path (instanton) that brings the system from the stable endemic state to the extinction of the disease, 
is given by the trajectory that minimises the WKB action $\mathcal S_{0}$. 
The optimal path must be a zero-energy trajectory. 
It turns out that there is no trajectory going directly from $A$ to $C$ (see Fig.~\ref{SIextinction}), instead, the fluctuational extinction point $B$ is crucial in the disease extinction.

The mean extinction time of the disease  $\tau$ is exponentially large in $N \gg 1$ and

\begin{align}
\tau \sim \exp \{N\mathcal{S}_0[\text{optimal path}]\},
\end{align}

where 

\begin{align}
\label{effectiveSI}
{\mathcal S}_0[\text{optimal path}]=\int^{\infty}_{-\infty}  (p_{S}\dot{S}+p_{I}\dot{I}) \ dt,
\end{align}

and the integration is evaluated along the optimal path  going from $A$ to $C$ ~\citep{Alex2008pre}.

\color{black}
For populations of more than one species interacting with each other, the analytical form of the mean extinction time is not available ~\citep{WKBAssaf}, and the optimal path can be computed only numerically.
It is also worth mentioning that, for extinction from a deterministically stable limit cycle driven by large fluctuations, the corresponding mean extinction time is also exponentially large in the population size $N$~\citep{smith2016extinction}.

\subsection{Extinction from marginally stable equilibrium states}
If the extinction is not driven by rare events, it can occur much more quickly. 
As we will see, the mean extinction time may have a power-law dependence on the population size in the predator-prey and competitive Lotka-Volterra models.
In these models, since extinction is driven by Gaussian fluctuations, the Fokker-Planck approximation can be applied.

We first take the classic Lotka-Volterra predator-prey model as an example.
Use the continuous variables $q_1$ and $q_2$ to represent the predator and prey populations, the deterministic rate equations are:

\begin{align}
\begin{split}
\label{LV}
\frac{dq_1}{dt}&=-\sigma q_1  + \lambda q_1 q_2,   \\
\frac{dq_2}{dt}&=\mu q_2 -\lambda q_1 q_2,
\end{split}
\end{align}

where $\sigma$ represents the death rate of the predator, $\mu$ represents the birth rate of the prey, and $\lambda$ is the rate of interaction between a predator and a prey.
Note that this formulation assumes that the preys have no intrinsic death, their population will grow exponentially without the presence of the predator.
There are three fixed points: $(q_1, q_2)=(0, 0)$, $(0,\infty)$, and $(\mu/\lambda,\sigma/\lambda)$. 
The first one corresponds to the case
where both species are extinct. 
The second one describes the population explosion of the prey due to the extinction of the predator. 
The third one represents the steady state where the predator and the prey coexist at the population size $N_1=\mu/\lambda$ and $N_2=\sigma/\lambda$, respectively.

\begin{figure}[!t]
	\center
	\includegraphics[width=3.5in]{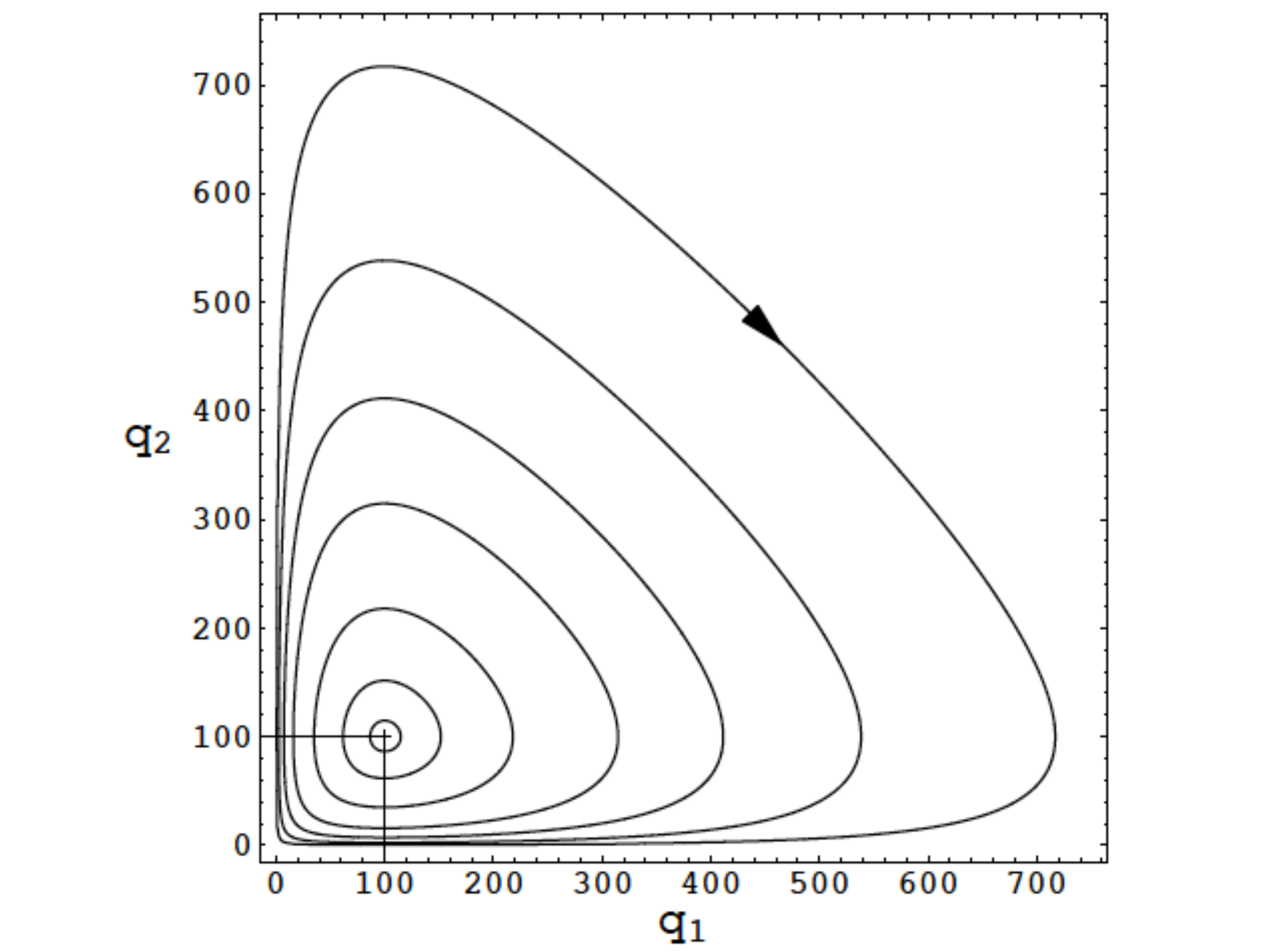}
	\caption{Orbits of constant $G=(0.01, 0.1, 0.4, 1, 1.7, 2.7, 4.2)$ in units of $\sqrt{\sigma \mu}$. The evolution proceeds clockwise around the mean-field fixed point of $N_1=N_2=100$. ~\citep{ parker:PRE:2009}. Permission for reused obtained from the publisher.} 
	\label{LVGF}
\end{figure}

A particular feature of the Lotka-Volterra model is that there is an ``accidental" conserved quantity: 

\begin{align}
G=\lambda q_1 -\mu -\mu \ln (q_1 \lambda /\mu)+\lambda q_2-\sigma -\sigma \ln (q_2 \lambda/\sigma),
\end{align}

where $G=0$ corresponds to the coexistence fixed point, and $G>0$ corresponds to larger amplitude cycles~\citep{parker:PRE:2009}. 
An illustration of orbits at different $G$ values is shown in Fig.~\ref{LVGF}. 
For a given initial condition, the the predator and prey populations cycle along a closed orbit. 

The existence of an ``accidental" conserved quantity $G$ not only leads to closed orbits, but also makes them marginally stable.
Population fluctuations due to demographic noise are isotropic in the space $(q_1,q_2)$, leading to slow diffusion between the mean-field orbits.
Even large deviations from a mean-field orbit,
such as extinction, can be seen as the accumulation
of many small step fluctuations in the radial direction.
This is in contrast with the systems with a stable fixed point or limit cycle, such as the endemic SIR model discussed in the previous section.
In those systems, large deviations proceed only along very special optimal paths in the phage space~\citep{DykmanJCP1994,elgart2004rare,Alex2008pre}. 
Consequently, the mean extinction time in marginally stable systems such as the predator-prey model has a power law dependence on the sizes of the two populations.

In the presence of demographic noises, the corresponding master equation is 

\begin{align}
\label{LVmaster}
\frac{dP(m,n,t)}{dt}&=\sigma \left[(m+1)(P(m+1,n)-mP(m,n))+\mu(n-1)P(m,n-1)-nP(m,n)\right]\nonumber\\
&+\mu\left[(n-1)P(m,n-1)-nP(m,n)\right]\nonumber\\ &+\lambda\left[(m-1)(n+1)P(m-1,n+1)-nmP(m,n)\right],
\end{align}

where $P(m,n,t)$ is the probability of the system having $m$ predators and $n$ preys at time $t$.

Since extinction in this case is driven by Gaussian fluctuations rather than large jumps, the Fokker-Planck approximation can be appropriately applied. $G$ can be identified as a ``slow" dynamic variable that is responsible for the long time behaviour of the system. 
In the presence of demographic stochasticity, after averaging out the ``fast" variable (angles in $(q_1,q_2)$ space), one can obtain a one-dimensional Fokker-Planck equation on the probability distribution of $G$. Solving the mean first passage time of this one-dimensional problem gives that the mean extinction time $\tau \sim N^{3/2}_1/N^{1/2}_2$ with $N_1\leq N_2$~\citep{parker:PRE:2009}.  

In the previous example of the predator-prey Lotka-Volterra model, overcrowding and intra-specific competition are not considered. 
The death of prey is solely caused by predation, and the per capita reproduction rate of predators only depends on the abundance of prey.
These paradise-like conditions are seldom met in real biological systems. 
Instead, competition is the norm and battles over resources for survival and reproduction can often be fierce and unforgiving. 
The competitive Lokta-Volterra model captures the self-limiting behaviour of the population growth. 
The corresponding deterministic rate equations are:

\begin{align}
\frac{dx}{dt}=r_1 x \left(1 - x-\alpha y\right),   \\
\frac{dy}{dt}=r_2 y \left(1 - y-\alpha x\right),
\end{align}

where $x=n_1/K_1$, $y=n_2/K_2$ are rescaled population size, in which $n_1$ and $n_2$ are the population size of each of the competing species, $K_{i}$ is the carrying capacity for each of them,
$r_1$ and $r_2$ are the intrinsic optimal growth rates of 
the two species when competition is absent, and $\alpha \in [0,1]$ is the competition
coefficient between the two species. 

In the limiting case when $\alpha = 0$, the growth of the two species are independent of each other.
When $0<\alpha<1$, there is an attracting fixed point $x^*_1=y^*_2=1/(1+\alpha)$ where the two species coexist.
If $\alpha = 1$, the two species are competitively identical.
Consdiering that they have the same carrying capacity $K_1=K_2=K$, the only difference is that one species reproduces faster and dies sooner than the other.
This leads to the degenerate case where there is a line of fixed points corresponding to the marginally stable coexistence of the two species with the ratio of populations determined uniquely by the initial conditions. 
In the degenerate case, the Fokker-Planck approximation can be applied. 
The corresponding Fokker-Planck equation is equivalent to stochastic differential equations of $x(t)$ and $y(t)$, which can be reduced to one-dimension by introducing $z(t)=x(t)-y(t)$:

\begin{align}
dz=v(z)+ \sqrt{2 D(z)} d W(t). 
\end{align}

Here $W(t)$ is a Wiener process. By determining the drift $v(z)$ and the diffusion $D(z)$ terms, the absorption time (the time until one of the species goes extinct) is $\tau \sim K$~\citep{lin:JSP:2012}. 

\citet{parsons:TPB:2008} also studied the competition dynamics of a fast-living species and a slow-living species, which have the same carrying capacity. 
The authors compared the absorption time to the prediction of the corresponding Wright-Fisher model of fixed population size, and found that it depends on the relative abundance of the two species. 
The absorption time is longer when the initial frequency of the fast-living species is higher, and shorter when it is lower. 
The work of \citep{kogan:PRE:2014} incorporated the "fast" and "slow" life history features with infectious diseases dynamics and studied the absorption time under the scenario of two pathogens competing for the same susceptible host population, in which one pathogen has higher infection rate yet its hosts recover more quickly compared the other pathogen.
Additional interesting works on extinction along a quasi-neutral line where population dynamics can be validly modelled by the Fokker-Planck approximation include \cite{parsons:TPB:2007} and \citet{constable:JPA:2013}.

\section{Discussion and conclusions}

In this paper, we provide a comparative analysis of the WKB and Fokker-Planck approximation methods in analysing the problem of population extinction under weak demographic fluctuations.  
In particular, we focus on estimating the mean extinction/absorption time of well-mixed systems containing a single or two interacting species.
The mean extinction time has distinct behaviours depending on the nature of the stationary states (fixed points) of the corresponding deterministic model.  
If the fixed point is attractive (for instance, logistic growth model and the endemic SIR model), the extinction is driven by rare events and the mean extinction time is experientially large in population size. 
In this case, the WKB method gives rise to the correct result whereas the Fokker-Planck approximation leads to an exponentially large error in the mean extinction time.
If the stationary state is marginally stable (for instance, the competitive Lotka-Volterra model when the two species have the same carrying capacity), the extinction instead is driven by typical Gaussian fluctuations and the mean extinction time has a power law dependence on the population size. 
Under this situation, the Fokker-Planck approach is also appropriate.

Here we only included examples of applying the WKB method in analysing a few basic population dynamics models, but note that the method has much broader applications in stochastic population dynamics. For instance, it provides a powerful tool in studying population extinction in fragmented landscape with dispersal between habitat patches \citep{meerson:PRE:2011, khasin:PRL:2012} and on heterogeneous networks \citep{hindes:PRL:2016, hindes:PRE:2017}. 
In addition, it has been applied to study the most likely path of extinction from species coexistence in the context of evolutionary games \citep{park:PRE:2017}.
For further reading on the vast applications of  the WKB approximation method, we recommend the following reviews and references therein. 
The concise review of \citet{ovaskainen:TREE:2010} provides an excellent overview of the WKB approximation in single species stochastic population models. 
Technique-wise, \citet{weber2017master} provides a comprehensive introduction to the path integral representation of master equations. 
The recent review of \citet{WKBAssaf} includes great details on applications of the WKB method in various models and pointed out interesting open questions.

Through this paper, 
we hope to arouse in biologists the interest to the WKB method and the great potential of applying it to solving stochastic population dynamics problems. 
Using several examples of the successful applications of the WKB and Fokker-Planck methods in solving evolutionary biology problems, we highlight the great value of knowledge transfer between physics and biology, and we encourage further exchange of knowledge and collaborations between physicists and biologists for developing novel approaches in modelling biological evolution.

\section*{Acknowledgement}

We thank Ping Ao, George Constable, Andrew Morozov, Ira B. Schwartz, Xiaomei Zhu, and an anonymous reviewer for useful discussions and/or comments.
Both authors are grateful to the Institute of Theoretical Physics of the Chinese Academy of Sciences, the collaboration was made possible through its Young Scientists' Forum on Theoretical Physics and Interdisciplinary Studies.

\bibliographystyle{plainnat}


\begin{thebibliography}{42}
\providecommand{\natexlab}[1]{#1}
\providecommand{\url}[1]{\texttt{#1}}
\expandafter\ifx\csname urlstyle\endcsname\relax
  \providecommand{\doi}[1]{doi: #1}\else
  \providecommand{\doi}{doi: \begingroup \urlstyle{rm}\Url}\fi

\bibitem[Assaf and Meerson(2010)]{Assafpre2010}
Michael Assaf and Baruch Meerson.
\newblock Extinction of metastable stochastic populations.
\newblock \emph{Physical Review E}, 81:\penalty0 021116, 2010.

\bibitem[Assaf and Meerson(2017)]{WKBAssaf}
Michael Assaf and Baruch Meerson.
\newblock W{KB} theory of large deviations in stochastic populations.
\newblock \emph{Journal of Physics A: Mathematical and Theoretical},
  50:\penalty0 263001, 2017.

\bibitem[Black et~al.(2012)Black, Traulsen, and Galla]{Black2012prl}
Andrew~J. Black, Arne Traulsen, and Tobias Galla.
\newblock Mixing times in evolutionary game dynamics.
\newblock \emph{Phys. Rev. Lett.}, 109:\penalty0 028101, 2012.

\bibitem[Bressloff and Newby(2014)]{bressloff2014path}
Paul~C Bressloff and Jay~M Newby.
\newblock Path integrals and large deviations in stochastic hybrid systems.
\newblock \emph{Physical Review E}, 89\penalty0 (4):\penalty0 042701, 2014.

\bibitem[Brillouin(1926)]{brillouin1926}
L{\'e}on Brillouin.
\newblock La m{\'e}canique ondulatoire de schr{\"o}dinger; une m{\'e}thode
  g{\'e}n{\'e}rale de r{\'e}solution par approximations successives.
\newblock \emph{CR Acad. Sci}, 183:\penalty0 24--26, 1926.

\bibitem[Chen et~al.(2017)Chen, Huang, Zhang, and Li]{Chen2017JSM}
Hanshuang Chen, Feng Huang, Haifeng Zhang, and Guofeng Li.
\newblock Epidemic extinction in a generalized susceptible-infected-susceptible
  model.
\newblock \emph{Journal of Statistical Mechanics: Theory and Experiment},
  2017:\penalty0 013204, 2017.

\bibitem[Constable et~al.(2013)Constable, McKane, and
  Rogers]{constable:JPA:2013}
George~WA Constable, Alan~J McKane, and Tim Rogers.
\newblock Stochastic dynamics on slow manifolds.
\newblock \emph{Journal of Physics A: Mathematical and Theoretical},
  46:\penalty0 295002, 2013.

\bibitem[Doering et~al.(2005)Doering, Sargsyan, and Sander]{doering:MMS:2005}
Charles~R Doering, Khachik~V Sargsyan, and Leonard~M Sander.
\newblock Extinction times for birth-death processes: Exact results, continuum
  asymptotics, and the failure of the fokker--planck approximation.
\newblock \emph{Multiscale Modeling \& Simulation}, 3:\penalty0 283--299, 2005.

\bibitem[Dykman et~al.(1994)Dykman, Mori, Ross, and Hunt]{DykmanJCP1994}
M.~I. Dykman, Eugenia Mori, John Ross, and P.~M. Hunt.
\newblock Large fluctuations and optimal paths in chemical kinetics.
\newblock \emph{The Journal of Chemical Physics}, 100:\penalty0 5735--5750,
  1994.

\bibitem[Elgart and Kamenev(2004)]{elgart2004rare}
Vlad Elgart and Alex Kamenev.
\newblock Rare event statistics in reaction-diffusion systems.
\newblock \emph{Physical Review E}, 70\penalty0 (4):\penalty0 041106, 2004.

\bibitem[Ewens(2004)]{ewens:book:2004}
W.~J. Ewens.
\newblock \emph{Mathematical Population Genetics. I. Theoretical Introduction}.
\newblock Springer, New York, 2004.

\bibitem[Fisher(1922)]{fisher:PRSE:1922}
R.~A. Fisher.
\newblock On the dominance ratio.
\newblock \emph{Proceedings of the Royal Society of Edinburgh}, 42:\penalty0
  321--341, 1922.

\bibitem[Gardiner(1985)]{gardiner1985handbook}
CW~Gardiner.
\newblock Handbook of stochastic methods.
\newblock \emph{Springer-Verlag, Berlin}, 1985.

\bibitem[Haefner(2012)]{haefner:book:2012}
James~W Haefner.
\newblock \emph{Modeling biological systems: principles and applications}.
\newblock Springer Science \& Business Media, 2012.

\bibitem[Hindes and Schwartz(2016)]{hindes:PRL:2016}
Jason Hindes and Ira~B Schwartz.
\newblock Epidemic extinction and control in heterogeneous networks.
\newblock \emph{Physical Review Letters}, 117:\penalty0 028302, 2016.

\bibitem[Hindes and Schwartz(2017)]{hindes:PRE:2017}
Jason Hindes and Ira~B Schwartz.
\newblock Epidemic extinction paths in complex networks.
\newblock \emph{Physical Review E}, 95:\penalty0 052317, 2017.

\bibitem[Kamenev and Meerson(2008)]{Alex2008pre}
Alex Kamenev and Baruch Meerson.
\newblock Extinction of an infectious disease: A large fluctuation in a
  nonequilibrium system.
\newblock \emph{Phys. Rev. E}, 77:\penalty0 061107, 2008.

\bibitem[Kendall(1966)]{kendall:JMS:1966}
David~G Kendall.
\newblock Branching processes since 1873.
\newblock \emph{Journal of the London Methematical Society}, 1:\penalty0
  385--406, 1966.

\bibitem[Khasin et~al.(2012)Khasin, Meerson, Khain, and
  Sander]{khasin:PRL:2012}
Michael Khasin, Baruch Meerson, Evgeniy Khain, and Leonard~M Sander.
\newblock Minimizing the population extinction risk by migration.
\newblock \emph{Physical Review Letters}, 109:\penalty0 138104, 2012.

\bibitem[Kimura(1964)]{kimura:JAP:1964}
Motoo Kimura.
\newblock Diffusion models in population genetics.
\newblock \emph{Journal of Applied Probability}, 1:\penalty0 177--232, 1964.

\bibitem[Kogan et~al.(2014)Kogan, Khasin, Meerson, Schneider, and
  Myers]{kogan:PRE:2014}
Oleg Kogan, Michael Khasin, Baruch Meerson, David Schneider, and Christopher~R
  Myers.
\newblock Two-strain competition in quasineutral stochastic disease dynamics.
\newblock \emph{Physical Review E}, 90:\penalty0 042149, 2014.

\bibitem[Kramers(1926)]{kramers1926}
Hendrik~A Kramers.
\newblock Wellenmechanik und halbzahlige quantisierung.
\newblock \emph{Zeitschrift f{\"u}r Physik A Hadrons and Nuclei}, 39:\penalty0
  828--840, 1926.

\bibitem[Landau and Lifshitz(2013)]{FermiGolden}
Lev~Davidovich Landau and Evgenii~Mikhailovich Lifshitz.
\newblock \emph{Quantum mechanics: non-relativistic theory}, volume~3.
\newblock Elsevier, 2013.

\bibitem[Lin et~al.(2012)Lin, Kim, and Doering]{lin:JSP:2012}
Yen~Ting Lin, Hyejin Kim, and Charles~R Doering.
\newblock Features of fast living: on the weak selection for longevity in
  degenerate birth-death processes.
\newblock \emph{Journal of Statistical Physics}, 148:\penalty0 647--663, 2012.

\bibitem[McElreath and Boyd(2008)]{mcelreath:book:2008}
Richard McElreath and Robert Boyd.
\newblock \emph{Mathematical models of social evolution: A guide for the
  perplexed}.
\newblock University of Chicago Press, 2008.

\bibitem[Meerson and Sasorov(2011)]{meerson:PRE:2011}
Baruch Meerson and Pavel~V Sasorov.
\newblock Extinction rates of established spatial population.
\newblock \emph{Physical Review E}, 83:\penalty0 011129, 2011.

\bibitem[Murray(2007)]{murray:book:2007}
J.~D. Murray.
\newblock \emph{Mathematical Biology I: An Introduction}.
\newblock Springer, 3rd edition, 2007.

\bibitem[Ovaskainen and Meerson(2010)]{ovaskainen:TREE:2010}
Otso Ovaskainen and Baruch Meerson.
\newblock Stochastic models of population extinction.
\newblock \emph{Trends in Ecology \& Evolution}, 25:\penalty0 643--652, 2010.

\bibitem[Park and Traulsen(2017)]{park:PRE:2017}
Hye~Jin Park and Arne Traulsen.
\newblock Extinction dynamics from metastable coexistences in an evolutionary
  game.
\newblock \emph{Physical Review E}, 96:\penalty0 042412, 2017.

\bibitem[Parker and Kamenev(2009)]{parker:PRE:2009}
Matthew Parker and Alex Kamenev.
\newblock Extinction in the lotka-volterra model.
\newblock \emph{Physical Review E}, 80:\penalty0 021129, 2009.

\bibitem[Parsons and Quince(2007)]{parsons:TPB:2007}
Todd~L Parsons and Christopher Quince.
\newblock Fixation in haploid populations exhibiting density dependence i: the
  non-neutral case.
\newblock \emph{Theoretical Population Biology}, 72:\penalty0 121--135, 2007.

\bibitem[Parsons et~al.(2008)Parsons, Quince, and Plotkin]{parsons:TPB:2008}
Todd~L Parsons, Christopher Quince, and Joshua~B Plotkin.
\newblock Absorption and fixation times for neutral and quasi-neutral
  populations with density dependence.
\newblock \emph{Theoretical Population Biology}, 74:\penalty0 302--310, 2008.

\bibitem[Risken(1996)]{risken1996fokker}
Hannes Risken.
\newblock Fokker-planck equation.
\newblock pages 63--95, 1996.

\bibitem[Shaffer(1981)]{shaffer:BS:1981}
Mark~L Shaffer.
\newblock Minimum population sizes for species conservation.
\newblock \emph{BioScience}, 31:\penalty0 131--134, 1981.

\bibitem[Smith and Meerson(2016)]{smith2016extinction}
Naftali~R Smith and Baruch Meerson.
\newblock Extinction of oscillating populations.
\newblock \emph{Physical Review E}, 93\penalty0 (3):\penalty0 032109, 2016.

\bibitem[Svirezhev and Passekov(2012)]{Svirezhev:book:2012}
Yuri~M Svirezhev and Vladimir~P Passekov.
\newblock \emph{Fundamentals of mathematical evolutionary genetics}.
\newblock Springer Science \& Business Media, 2012.

\bibitem[Touchette(2009)]{touchette2009large}
Hugo Touchette.
\newblock The large deviation approach to statistical mechanics.
\newblock \emph{Physics Reports}, 478\penalty0 (1):\penalty0 1--69, 2009.

\bibitem[Traill et~al.(2007)Traill, Bradshaw, and Brook]{traill:BC:2007}
Lochran~W Traill, Corey~JA Bradshaw, and Barry~W Brook.
\newblock Minimum viable population size: a meta-analysis of 30 years of
  published estimates.
\newblock \emph{Biological Conservation}, 139:\penalty0 159--166, 2007.

\bibitem[Tsoularis and Wallace(2002)]{tsoularis:MB:2002}
A~Tsoularis and J~Wallace.
\newblock Analysis of logistic growth models.
\newblock \emph{Mathematical Biosciences}, 179\penalty0 (1):\penalty0 21--55,
  2002.

\bibitem[Verhulst(1838)]{verhulst:Quetelet:1838}
Pierre-Fran{\c{c}}ois Verhulst.
\newblock Notice sur la loi que la population suit dans son accroissement.
\newblock \emph{Correspondance Math{\'e}matique et Physique Publi{\'e}e par A.
  Quetelet}, 10:\penalty0 113--121, 1838.

\bibitem[Weber and Frey(2017)]{weber2017master}
Markus~F Weber and Erwin Frey.
\newblock Master equations and the theory of stochastic path integrals.
\newblock \emph{Reports on Progress in Physics}, 80\penalty0 (4):\penalty0
  046601, 2017.

\bibitem[Wentzel(1926)]{wentzel1926}
Gregor Wentzel.
\newblock Eine verallgemeinerung der quantenbedingungen f{\"u}r die zwecke der
  wellenmechanik.
\newblock \emph{Zeitschrift f{\"u}r Physik A Hadrons and Nuclei}, 38:\penalty0
  518--529, 1926.

\end{thebibliography}

\end{document}